\documentclass[%
 aip,
 amsmath,amssymb,
 reprint,%
]{article}

\usepackage{graphicx}
\usepackage{dcolumn}
\usepackage{bm}
\usepackage{here}
\usepackage[utf8]{inputenc}
\usepackage[T1]{fontenc}
\usepackage{mathptmx}
\usepackage{etoolbox}
\usepackage{diagbox}
\begin{document}


\title{Remote and Noncontact Acceleration and Suppression of Heat Convection on Object Surfaces using Intensity-Adjusted Acoustic Streaming}
\author{Yuki Yuzawa, Keisuke Hasegawa}%

\date{\today}

\begin{abstract}
The temperature of objects in an environment is often controlled by accelerating or suppressing heat convection.
Conventional methods generally involve insulation for heat retention and forced convection for cooling. 
However, there is no practical method for remotely and target-selectively controlling heat convection in objects in room environments.
In this study, we demonstrated that intensity-adjusted acoustic streaming—ultrasound-driven airflow—can both suppress natural convection and induce forced convection on the surface of a heat source.
The direct acceleration and deceleration of air over the object surface were induced by an ultrasonic beam with high temporal and spatial resolution.
The experiment showed that it is possible to switch between insulating and cooling by adjusting the ultrasonic output intensity, and that it is essential for heat retention to prevent heat from escaping from the object surface by convection using acoustic streaming.
We consider these results to lead to the realization of a situation where virtual heat-retaining blankets or cooling fans that are customized for specific objects in the environment can be ubiquitously attached to them, opening the door to applications such as temperature control of food and beverages and remote control of temperature profiles in manufacturing processes and laboratory settings.
 
\end{abstract}

\maketitle

Objects exchange heat with the surrounding medium via heat convection or heat conduction and finally reach thermal equilibrium with the environmental temperature\cite{ref1}\cite{ref2}.
When controlling the object temperature, there are many cases where it is enough to promote the natural cooling of objects or to delay reaching thermal equilibrium between objects and the surrounding air.
Examples of the former include heat sinks\cite{ref3}\cite{ref4}, air cooling apparatuses\cite{ref5}\cite{ref6}, and liquid cooling\cite{ref7} for electronic devices such as computers.
These are often used in combination with whole-building heat exchange \cite{ref8}.
Examples of the latter include the use of heat-insulating containers and clothing to adjust the temperature of food and humans\cite{ref9}\cite{ref10}\cite{ref11}.
On a larger spatial scale, examples include insulated walls and windows designed to improve the effect of indoor air conditioning\cite{ref12}\cite{ref13}\cite{ref14}\cite{ref15}.
These methods can essentially be described as "contact" methods because components such as heaters, fans, or insulators must be placed near the heat source.

Besides the case with those methods, if remote cooling and insulation could be achieved, applications that differ from the current ones would be realized.
For example, an additional heat-transfer path for a mobile computer, separate from the built-in fan, would allow more demanding calculations without worrying about thermal throttling.
This is often important for tablet computers without fans, where large heat sinks are difficult to install\cite{ref16}\cite{ref17}.
Other potential applications include remotely maintaining the temperature of food in portable containers, such as coffee in paper cups\cite{ref18} and meals served in banquet halls.  
However, there are only a few examples of prior research related to them.
Examples include localized heating using a halogen lamp with an electronically controlled mirror\cite{ref19} and localized cooling by airborne mist on skin surfaces transported via ultrasound-induced acoustic streaming\cite{ref20}.
These are designed to present thermal or cooling stimuli to the human body surface and are not intended to induce significant temperature changes.

\begin{figure}[h]
    \centering
    \includegraphics[width=\columnwidth]{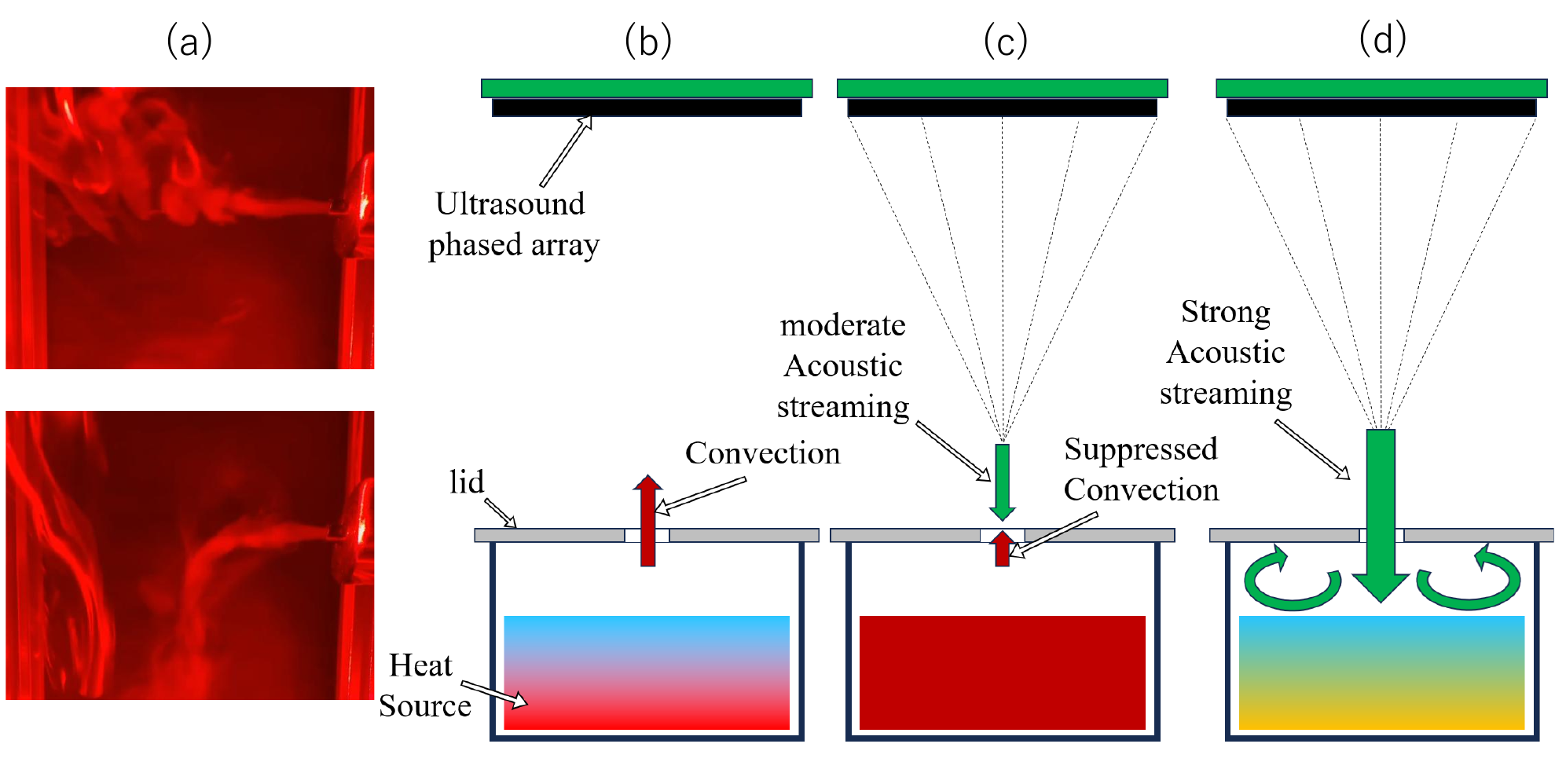}
    \caption{(a) Visualized acoustic streaming: floating smoke (upper figure) is guided downward by the acoustic streaming (lower figure).
    (b) Concept of the application scenario, where the temperature-control target in an open container is placed under the ultrasound phased array.
    (c) In heat retention, moderate acoustic streaming suppresses convection over the target.
    (d) In heat dissipation (cooling), strong acoustic streaming causes forced convection to expedite cooling of the target.}
    \label{fig1}
\end{figure}

Acoustic streaming differs from air control by fans or jets in that it directly accelerates fluid within acoustic intensity fields generated by emitted ultrasonic beams.
This means that virtual jet outlets can be positioned at a distance from the acoustic radiation device, allowing for the selective cooling of specific targets.
Additionally, medium acceleration is known to be proportional to the propagating ultrasonic wave power\cite{ref21}, meaning that precise acceleration magnitude control of the air at the target surface can be realized by appropriate tuning of ultrasound emission intensity.
Based on these physical principles, we propose a method for remote heat convection control on the surface of a general target object using acoustic flow generated by an ultrasonic beam.
When the goal is cooling, high ultrasound power is used to induce forced convection, thereby promoting heat transfer of the target.
When the goal is heat retention, the ultrasound power is finely adjusted to generate an acoustic flow, which precisely suppresses natural convection at the target surface to inhibit heat transfer.
The cooling/heat-retaining modes are switched simply by adjusting the ultrasound power level (Fig.~\ref{fig1}).
Thus, acoustic streaming can serve both as an accelerator that generates airflow and as a brake that suppresses airflow that has already been generated. 
Nonlinear ultrasonics have been used to enhance heat transfer \cite{Li2025}, but there are no experimental precedents for suppressing existing convection using macroscopic acoustic streaming.
This paper presents the results of experiments to investigate the extent to which this method can accelerate or suppress temperature changes compared to the case of natural convection.
In addition to temperature changes, we visualized convection near the target surfaces using particle image velocimetry (PIV).
We conducted these experiments while varying the intensity of the ultrasound power and discussed the phenomenon by correlating the temperature changes in each case with the resulting convection patterns. 

Intense sound waves follow the nonlinear wave equation and exhibit nonlinear effects as they propagate through the air, resulting in time-constant components in the sound pressure and particle velocity.
This phenomenon is known as acoustic streaming\cite{ref22}\cite{ref23}\cite{ref24}\cite{ref25}, and this study focused on Eckart acoustic streaming, ultrasound-driven flows longer than the wavelength.
The Eckart acoustic streaming is often modeled as an incompressible flow, and it is assumed that its volumetric driving force $\bm{F}$ is determined by the acoustic intensity $\bm{I}$ by the following equation\cite{ref21}: 
\begin{equation}
\bm{F}=\frac{2\alpha}{\rho_0c_0}\bm{I}.
\end{equation}
Here, $\alpha, \rho_0, c_0$ represent the sound absorption coefficient of air, the medium density, and the speed of small-amplitude sound, respectively.
This equation indicates that the volumetric driving force generated by the ultrasound beam accelerates or decelerates the air within that region.
This force is expected to promote or suppress natural thermal convection of the target. 

\begin{figure}[h]
    \centering
    \includegraphics[width=\columnwidth]{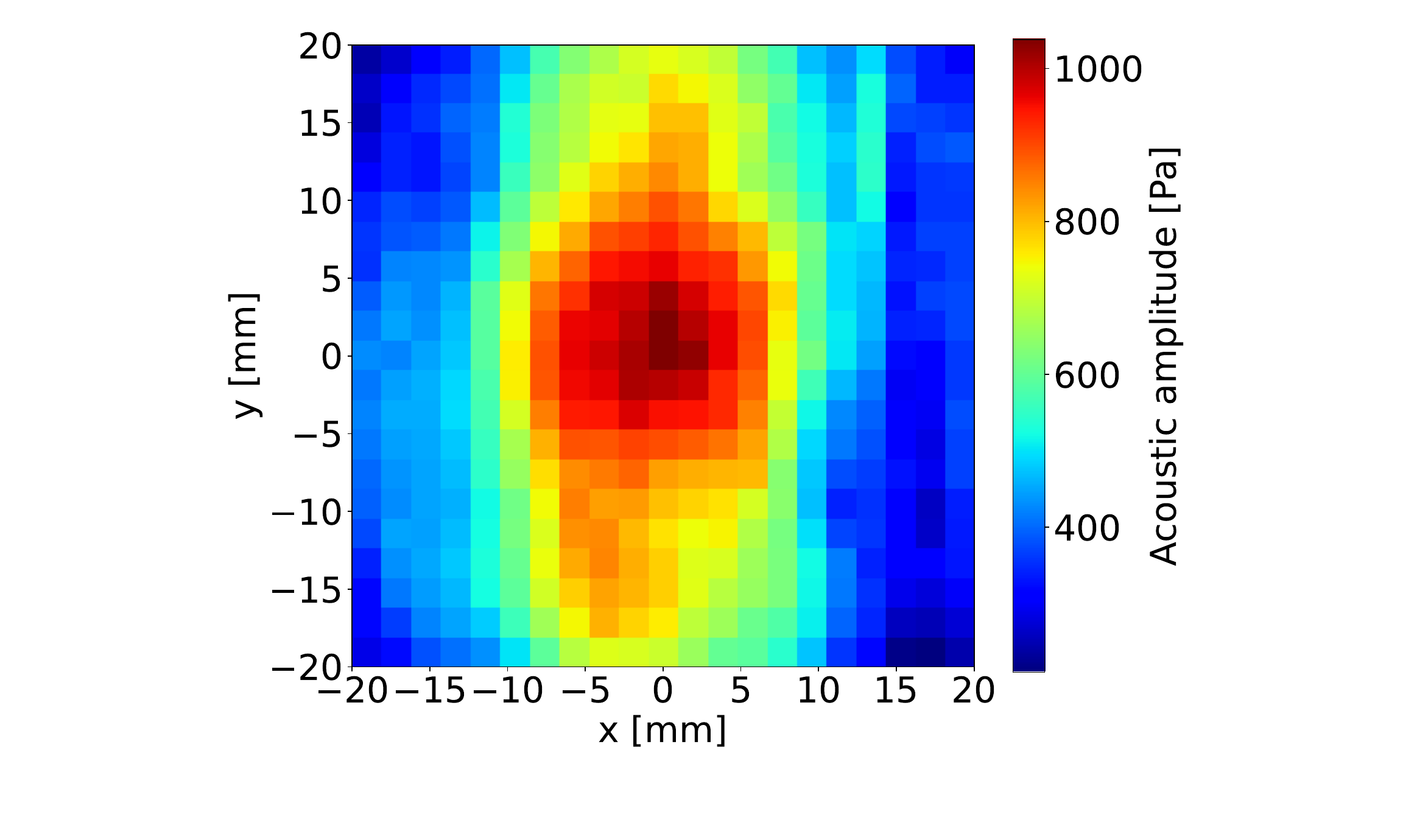}
    \caption{Measured sound pressure distribution around an ultrasound focus.}
    
    \label{fig2}
\end{figure}

In this study, a phased array of airborne ultrasound transducers\cite{ref26}. This is a collective ultrasound source consisting of many ultrasound transducers placed in a two-dimensional lattice pattern.
Each transducer is driven at the same frequency, with its phase and amplitude individually controlled to form an ultrasonic focus at desired spatial positions.
A concentrated acoustic intensity field is formed around the focal region. This region is generally stretching toward the normal of the emission plane of the array, and acoustic streaming is generated along that direction\cite{ref24}\cite{ref27}\cite{ref28}. 
The phased array was driven to generate acoustic streaming near the surface of a heat source.
The volumetric driving force of acoustic streaming was adjusted by electronically controlling the output of the phased array. 
Phased arrays can generate a wide variety of spatial distributions of ultrasound fields\cite{ref29}.
Among them, in this study, we generated an ultrasound focus since it can produce the possible maximum acoustic intensity.
This is achieved by applying phase compensation to each transducer's emission, proportional to the distance between the set focal position and it\cite{ref30}.
First, we measured the distribution of ultrasound pressure near the focus. 
The phased array was placed on a table, and the focus was formed at a height of 233~mm away from the transducer surface.
A standard microphone system  (1/8-in. microphone, type 4138-A-015; pre-amplifier, type 2670; condition amplifier, type 2690-A; all products of Hottinger, Br\"{u}el \& Kj\ae r, Germany) was positioned 315~mm away from the transducer surfaces and scanned across a horizontal plane using a two-dimensional linear actuator.
To prevent microphone overload, the measurements were taken with the phased array output adjusted.
Fig.~\ref{fig2} shows an example of measured ultrasound magnitudes around the focal position.

\begin{figure}[h]
    \centering
    \includegraphics[width=\columnwidth]{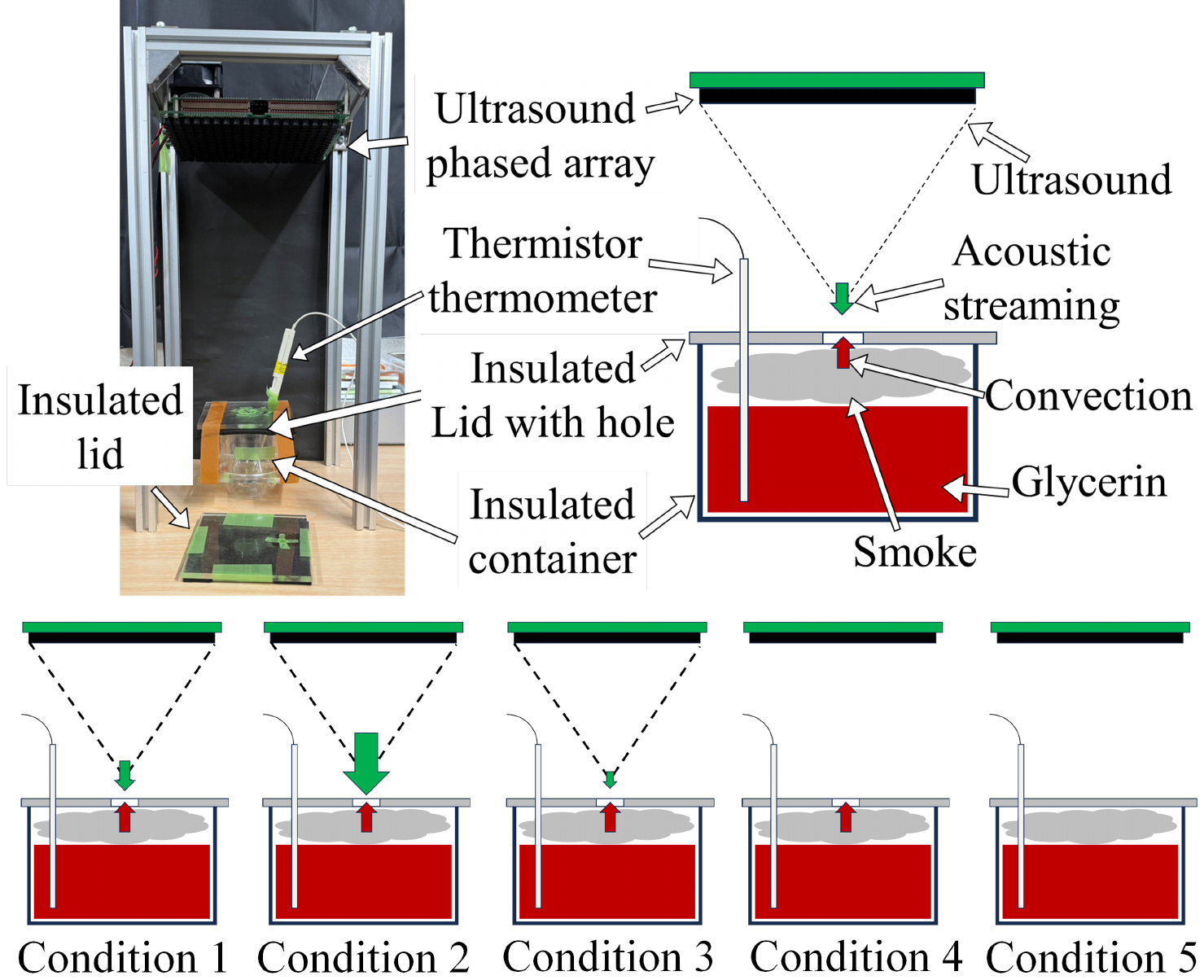}
    \caption{Experimental setup (upper figures) and conditions (lower figures).}
    
    \label{fig4}
\end{figure}

We then conducted experiments to quantitatively estimate the heat transfer control performance of the proposed method.
We measured temperature changes and observed the flow velocity field over the target using Particle Image Velocimetry (PIV). 
Fig.~\ref{fig4} shows a photograph and a schematic diagram of the experimental setup.
Schematic diagrams of each experimental condition are also shown here.
In the experiment, an ultrasound phased array module was mounted on an aluminum frame with its emission plane facing downward.
To improve the visibility of smoke inside the container, a blackout curtain was attached to the aluminum frame.
In the experiment, heated glycerin enclosed in an insulated container was used as the heat source.
It prevented fogging inside the container and mass loss due to vaporization during the experiment, making it suitable for observing the gas inside.

An acrylic lid was placed over the container. 
We used two types of lids: one had a hole with a diameter of 20~mm, and the other had no holes.
Insulation was applied to each lid to reduce heat transfer other than convection, namely thermal conduction and thermal radiation.
This ensured that heat exchange with the external gas occurred, as much as possible, solely through thermal convection via the hole.
Additionally, small holes for inserting thermometers were provided in all lids.
The container was made of glass to ensure good visibility during PIV measurements. 

In this setup, experiments were conducted with a heat source placed beneath the phased array.
The distance from the phased array to the hole in the lid was set to 315 mm. 
The focal depth from the emission plane of the array was set to 233~mm.
Via preliminary experiments, we found that this spatial configuration effectively suppressed gas leakage from the hole by acoustic streaming with tuned output intensity.

The experimental procedure is described below.
First, glycerin was heated to approximately 110~$^\circ$C and then poured into the container.
Next, smoke generated by a smoke generator was poured into the container.
By irradiating the system with a 638-nm laser sheet (KLD-V-U, Kato Koken, Japan) under these conditions, we visualized the gas flow inside the container and near the hole in the vertical plane.
The visualization images were captured using a high-speed camera (High-Speed Camera K5, Kato Koken, Japan) set to a frame rate of 250 fps.
The air conditioning system was kept running throughout the experiment.
The PIV analysis under each experimental condition was conducted using a one-second clip of the captured footage with good visibility.
Note that the region above the lid was not filled with smoke, and the PIV results in that region do not fully reflect the flow field; only whether smoke leaked from the hole was evaluated.
In the experiment, the point at which the glycerin temperature dropped to 90~$^\circ$C was set as the measurement onset time.
Temperature measurements were taken at one-minute intervals from that point until 10 minutes later, and the gas flow was observed.
The effect of acoustic streaming on heat transfer was evaluated by comparing temperature changes under various conditions.
The following five experimental conditions were set, and five measurements were taken for each condition. Note that the following pressure values are calculated by rescaling the measured focal pressure shown in \ref{fig2} based on additional acoustic measurements with varied output amplitudes: 

(1) A perforated lid was used, and the focused ultrasound pressure was set to 712 Pa (the sound pressure determined through preliminary experiments to effectively minimize gas migration outside the container)

(2) A perforated lid was used, and the focused ultrasound pressure was set to 1.18 kPa (the sound pressure determined through preliminary experiments to induce forced convection within the container)

(3) A perforated lid was used, and the focused ultrasound pressure was set to 276 Pa (a slightly lower sound pressure than in (1), determined through preliminary experiments to be the level that minimizes agitation of the liquid inside the container)

(4) A perforated lid was used, and no acoustic streaming was generated (corresponding to the control condition)

(5) A lid without holes was used, and no acoustic streaming was generated (minimum heat transfer between the inside and outside of the container is expected)

\begin{figure}[t]
    \centering
    \includegraphics[width=0.7\textwidth]{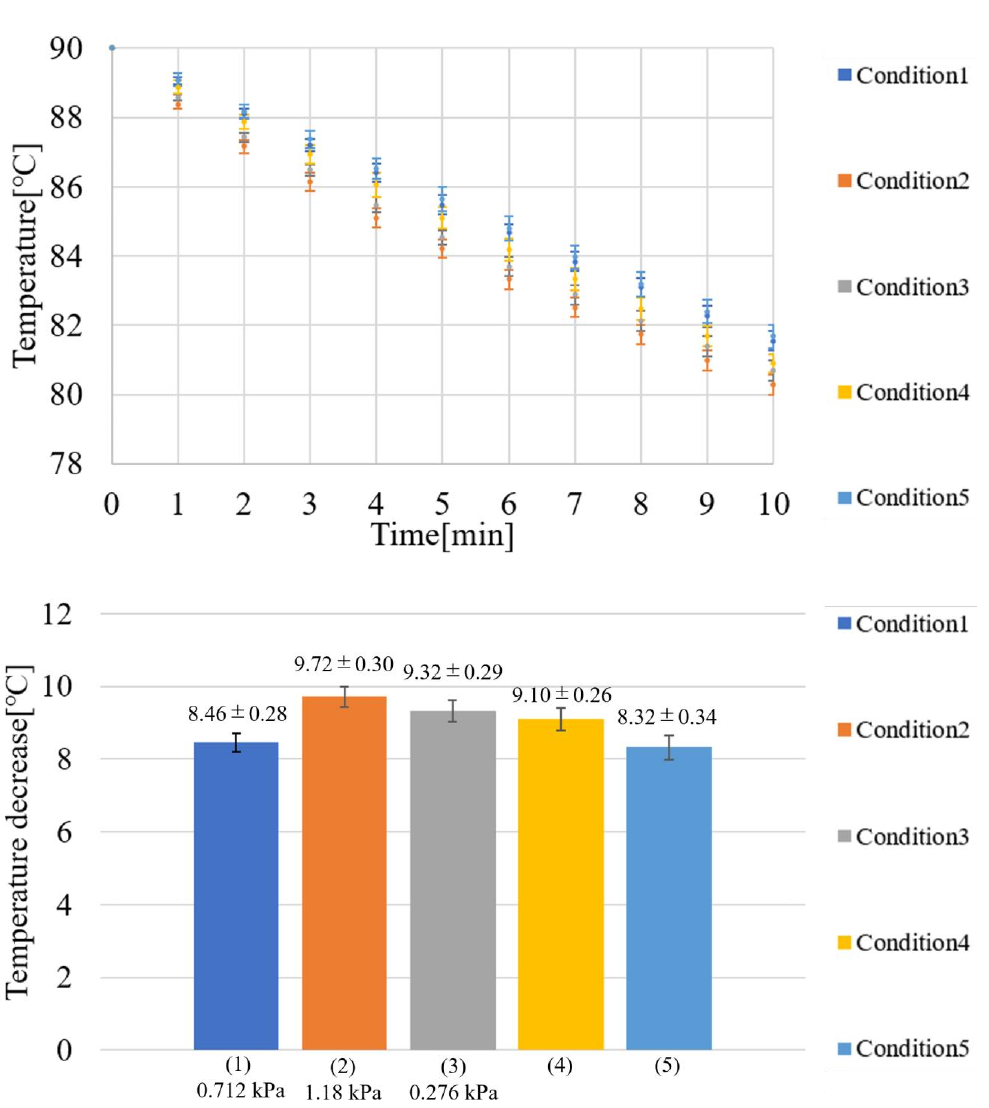}
    \caption{Average temporal profiles of glycerin temperature decreases under each condition. Error bars indicate standard deviations.}
    \label{fig5}
\end{figure}

Fig.~\ref{fig5} shows the average values and standard deviations of the results from five measurements of temperature changes under each experimental condition.
Additionally, their final values after 10 minutes under each condition are shown.
Table.~\ref{tab1} shows whether there were significant differences between the various conditions based on the results of the Mann-Whitney U test across each condition pair with a significance level of 0.05. 

\begin{table}
\caption{Results of the Mann–Whitney U test comparing temperature changes between experimental conditions. * indicates condition pairs between which differences are confirmed with a significance level of 0.05.}
\label{tab1}
\begin{tabular}{cccccc}
& Cond.1 & Cond.2 & Cond.3 & Cond.4 & Cond.5 \\
\hline
Condition1 & & * & * & * & - \\
Condition2 & * & & - & * & * \\
Condition3 & * & - & & - & * \\
Condition4 & * & * & - & & * \\
Condition5 & - & * & * & * & \\
\end{tabular}
\end{table}

Among the results, the most notable point is that no significant difference was observed in the state after 10 minutes when comparing conditions (1) and (5).
In other words, this demonstrates that our method achieved heat retention to the same degree as using a lid without holes.
Next, when comparing Condition (2) and (4), we find that forced convection through a small aperture significantly reduces the target temperature.
Note that, since the difference in temperature change was relatively small in the experiments using glycerin, a similar experiment using water as the temperature control target was also conducted.
The range of temperature change was greater than that observed with glycerin, and when comparing the temperature drop of water of 90$^\circ$C after 10 minutes, heat retention resulted in an average of 10.5 $\pm$ 0.28~$^\circ$C, and cooling resulted in an average of 14.1 $\pm$ 0.58~$^\circ$C. 
These results also demonstrate that an appropriately adjusted acoustic streaming provides a heat retention effect, while a strong acoustic streaming produces a cooling effect.

\begin{figure}
    \centering
    \includegraphics[width=0.7\columnwidth]{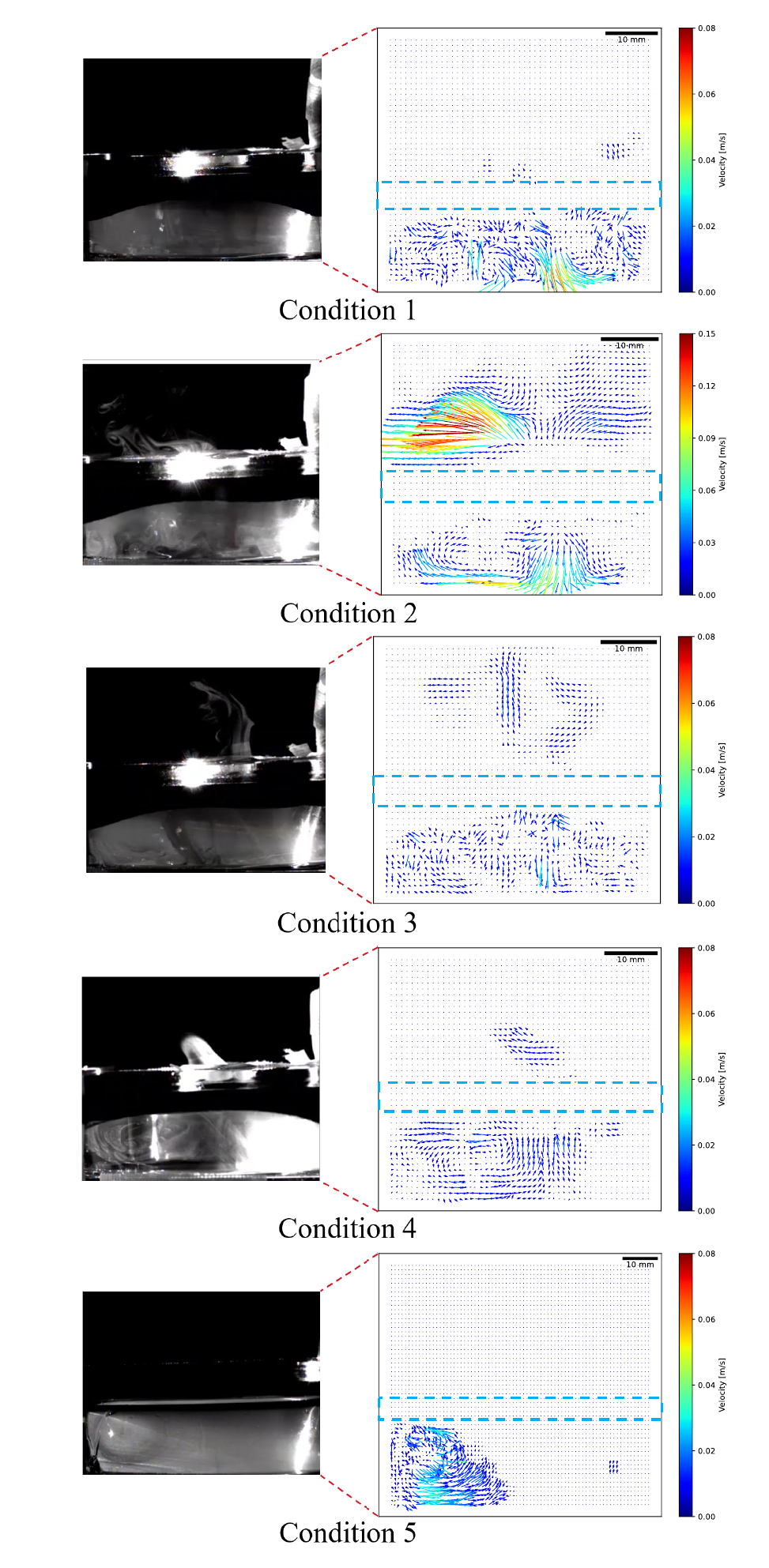}
    \caption{Gas flow under each condition. Blue dotted rectangles indicate the lid positions. }    
    \label{fig6}
\end{figure}

Fig.~\ref{fig6} shows the results of PIV measurements of the flow velocity fields inside the container and near the lid under each condition.
We see that under Condition (1), the smoke leakage from the opening to the outside of the container was largely suppressed, but at the liquid surface, smoke mixing due to convection was observed at the same time.
In contrast, while Condition (3) was able to suppress the movement of smoke at the liquid surface, it is shown that it did not prevent smoke from escaping through the holes.
A comparison of the heat loss effects reveals that Condition (1) is more effective than Condition (3).
Therefore, it was found that, for heat retention, it is more important to prevent gas from escaping from the container than to suppress the movement of gas at the liquid surface.
In light of this, if it were possible to simultaneously suppress the escape of hot air through the holes and suppress thermal convection at the liquid surface, a higher level of heat retention could potentially be achieved than with a simple lid alone, because the latter cannot be achieved with a simple lid.
Regarding condition (2), it was observed that forced convection caused by acoustic streaming disrupted natural convection and created waves on the liquid surface.
Also, smoke was escaping vigorously through the lid.
It is considered that this cooling effect would be even greater if forced convection caused by the acoustic streaming were directed at the liquid surface with the lid removed. 

On the other hand, the experiments revealed the following three issues with the proposed method. 
First, the size of the heat transfer surface to which this method can be applied is limited.
This is because of acoustic streaming generated by a single focus with a small cross-sectional area.
To address this problem, methods that expand the acoustic streaming's effective region through multi-focus formation\cite{ref29}\cite{ref31} or focus expansion\cite{ref32} are considered effective. 
Second, it is difficult to maintain the optimal conditions for suppressing the escape of hot air over a long period of time with a static ultrasound output due to gradual temperature changes in the target.
To address this problem, a feedback control system can be introduced that measures the temperature of the source and environment and dynamically adjusts the excitation amplitude in response to them. 
Third, the range within which the acoustic streaming directly acts on the gas inside the container was limited: the acoustic streaming did not directly act on the entire interior of the vessel.
As a result, it is possible that we were unable to fully evaluate the effect of internal mixing caused by the acoustic streaming on heat transfer.
To address this problem, acoustic streaming that acts over a wider area may be employed, in connection with the first problem. 

In this study, we proposed a method for achieving remote, non-contact heat transfer control by generating acoustic streaming by forming an ultrasound focus near the surface of a heat source, which imparts a volume-driving force to accelerate or decelerate convection. 
Based on the experimental results and PIV analysis, it was demonstrated that, with regard to heat retention, the acoustic streaming has two effects.
First, it suppresses the escape of hot air through the holes in the lid, and second, it suppresses thermal convection near the liquid surface. In particular, it was shown that, to achieve heat retention, it is more important to prevent the escape of hot air through the holes in the lid. Also, with regard to cooling, we demonstrated that acoustic streaming can cause a decrease in the temperature of the target surface by breaking the surface thermal convection and generating forced convection near the surface from the outside. 

As indicated in the above discussion, the primary focus of this study is on experimentally demonstrating the feasibility of suppressing or increasing heat transfer via acoustic streaming.
However, the experimental setup is highly artificial, and this does not guarantee the applicability of this method in various practical situations.
To address this problem, future research may focus on creating acoustic streaming covers over larger surfaces, as described above, or generating acoustic streaming fields that cover objects such as domes.
Also, real-time intensity control of the generated ultrasound field in response to temperature changes from moment to moment will be required for operations over a longer period.
By addressing these problems, it will finally be possible to build a physical system that can freely control heat exchange with the environment. For example, situations like attaching virtual blankets or cooling fans to various objects in the environment using ultrasound will be realized.

\bibliographystyle{amsplain}
\bibliography{aipsamp}

\end{document}